\documentclass[twocolumn,amsmath,floatfix, nofootinbib]{revtex4-1}
\usepackage{amssymb}
\usepackage{graphicx}
\usepackage{array}
\usepackage{hhline}
\usepackage{longtable}
\usepackage{bm}
\usepackage{hyperref}
\usepackage{color, soul}

\begin{document}

\newcommand{\rsp}[1]{\hspace{-0.15em}#1\hspace{-0.15em}}

\title{Chiral smectic $A$ membranes: Unified theory of free edge structure and twist walls}

\author{C. Nadir Kaplan$^{\ast}$\textit{$^{\dag}$} and Robert B. Meyer}
\affiliation{The Martin Fisher School
of Physics, Brandeis University, Waltham, Massachusetts 02454} 

\date{\today}

\begin{abstract}
Monodisperse suspensions of rodlike chiral $fd$ viruses are condensed into a rod-length thick colloidal monolayers of aligned rods by depletion forces. Twist deformations of the molecules are expelled to the monolayer edge as in a chiral smectic $A$ liquid crystal, and a cholesteric band forms at the edge. Coalescence of two such isolated membranes results in a twist wall sandwiched between two regions of aligned rods, dubbed $\pi$-walls. By modeling the membrane as a binary fluid of coexisting cholesteric and chiral smectic $A$ liquid-crystalline regions, we develop a unified theory of the $\pi$-walls and the monolayer edge. The mean-field analysis of our model yields the molecular tilt profiles, the local thickness change, and the crossover from smectic to cholesteric behavior at the monolayer edge and across the $\pi$-wall. Furthermore, we calculate the line tension associated with the formation of these interfaces. Our model offers insights regarding the stability and the detailed structure of the $\pi$-
wall and the monolayer edge.
\end{abstract}

\pacs{pacs} 
\maketitle

\footnotetext{$^{\ast}$~E-mail: \textit{nadir@seas.harvard.edu}}
\footnotetext{\dag~Present address: \textit{School of Engineering and Applied Sciences, Harvard University, Cambridge, Massachusetts 02138, USA.}}

\section{Introduction}

Chirality, the breaking of mirror symmetry, occurs at many length scales, from nanometer-sized DNA to the coiled form of Gastropod snails at various sizes~\cite{Snail1}. In soft matter, microscopic chirality can alter numerous macroscopic properties of materials. For example, in chiral smectic liquid crystals grain boundaries form above a critical temperature to alleviate the frustration between the layer formation and the twist tendency of the chiral molecules~\cite{Lubensky, TGB2, TGB3}. This is analogous to the penetration of magnetic field lines into superconductors by creating a lattice of parallel vortices~\cite{degennes3}. Based on this analogy, smectic $A$ samples maintain twist- and bend-free molecular orientations within each layer by the expulsion of these deformations to the edges or around isolated defects, similar to the well-known Meissner effect in superconductors~\cite{DeGennes2}.

One realization of this analogy is a model system of two dimensional colloidal membranes composed of aligned rods, which behave as a single layer of a chiral smectic A (Sm-$A^\ast$) liquid crystal. Monodisperse suspensions of the 880 nm long, rodlike wild-type $fd$ viruses are spontaneously condensed into these one rod length thick colloidal monolayers with free edges by attractive depletion forces~\cite{Barry2, Yang, depletion}. The penetration of twist at the monolayer edge could be directly observed, allowing for the confirmation of the de Gennes theory~\cite{Barry}. Recently, it was reported that the line tension of the monolayer edge, an important variable relevant to the formation energy of the free boundaries, can be controlled \textit{in situ} by the chirality of rods~\cite{gibaud}. Furthermore, it was found that monolayers which consist of achiral rods still possessed twist deformations at the edge. A curved edge profile was observed, which was robust for the entire range of measured line tension 
values.

By changing the line tension, colloidal Sm-$A^\ast$ membranes can easily be induced to exhibit morphological transitions~\cite{gibaud}. One way to achieve this is the coalescence of two or more membranes. When two membranes are surrounded by the edges with same handedness, the molecular twist localized at the edge poses a problem for the coalescence, as the direction of twist will be in opposite directions where the two edges meet. The final structure adopts a one-dimensional line defect between the two coalesced membranes when the sizes of the two membranes are comparable to each other. In order for the molecular arrangement across the defect to be continuous, the rods must twist by $\pi$ radians. Henceforth we will refer to these lines of twist as $\pi$-walls. These can also be imprinted in the monolayers using optical forces~\cite{Zakhary}. 

$\pi$-walls resemble various other phenomena observed in condensed matter systems. For instance, they are analogous to a laminar model of an array of alternating normal metal and superconducting regions, which was replaced by the Abrikosov flux-lattice phase in type-II materials~\cite{goodman, degennes3, sarma}. Furthermore, the pitch associated with the helicity of a cholesteric sample can be unwound by electric and magnetic fields, creating similar twist walls in the vicinity of the critical field~\cite{cholesteric1, cholesteric2}. Another example is magnetic systems, where the rotation of spin magnetic moments in space cause resembling domains called Bloch walls~\cite{Bloch}.

%In this paper we utilize the theory, first proposed in Ref.~\cite{Zakhary} for the $\pi$-walls, to extend the analysis to the monolayer edge.

In this paper we extend the analysis of the theory, first proposed in Ref.~\cite{Zakhary} for the $\pi$-walls, to the free edge of the colloidal monolayer. Our approach treats the monolayers as a binary liquid of Sm-$A^\ast$, where the rods are aligned parallel to the layer normal, and a cholesteric (chiral nematic -- $Ch$) region, which the Sm-$A^\ast$ order melts into both at the monolayer free edge and the $\pi$-wall. By introducing an order parameter field which is proportional to the local Sm-$A^\ast$ concentration, this framework couples the smectic order within the monolayer and the orientation of the nematic director field of the constituent molecules where the elastic distortions are pronounced. Our approach is reminiscent of the Landau-Ginzburg formalism of smectics put forward by de Gennes~\cite{DeGennes2}, although in the colloidal membranes there is no discrete translational symmetry as in a layered Sm-$A^\ast$ material. Thus, we replace the conventional smectic order parameter, 
which preserves the translational symmetry via a complex number, by a real order parameter in our analysis, which solely accounts for the local concentration of the regions of perfectly aligned rods within the monolayer. Henceforth, a chiral smectic $A$ monolayer, which lacks the mass-density wave of a layered smectic, shall be understood by the abbreviation ``Sm-$A^\ast$.'' In the present work, the effect of the depleting agent, \textit{i.e.} the nonadsorbing polymer, is modeled by appropriate surface tension terms, and plays a crucial role in determining the structure of the monolayer edge and the $\pi$-walls. We perform a mean-field analysis of our model to calculate the thickness, tilt angle, and the smectic order parameter profiles of the $\pi$-walls and the free edges. These profiles reveal a Sm-$A^\ast$-to-$Ch$ transition at both types of interfaces, as observed in experiments. We examine the effect of the dimensionless parameters in our model on the profiles, in order to determine the relevant parameters which induce pronounced changes in the results. Overall we find that our theory is robust within a reasonable range of all free parameters. This allows us to obtain the best fit of the tilt angle profile and the thickness to the experimental measurements at the free edge and the $\pi$-wall. Furthermore, we calculate the line tension of the membrane edge and the $\pi$-wall as a function of the molecular chirality, or equivalently temperature~\cite{gibaud}, in order to match them with experimental measurements. Accordingly, we investigate the thermodynamic stability of these structures with respect to each other.

The present study is organized as follows: In the next section, based on experimental evidence, we explain the hypothesized structures of the membrane edge and the $\pi$-wall (Sec.~\ref{subsec:Structure}), followed by a detailed description of our model (Sec.~\ref{subsec:FreeEnergy}). In Sec.~\ref{subsec:coalescence} we propose a simple analysis regarding the thermodynamic stability condition of two coalesced membranes with a $\pi$-wall inbetween with respect to two disconnected membranes. In Sec.~\ref{sec:Results} we present the results, where we apply our theory to the membrane edge and the $\pi$-wall in order to test our hypotheses. Following the discussion of methods in Sec.~\ref{subsec:Methods}, we present in Sec.~\ref{subsec:ParameterTuning} how the profiles are altered when the free parameters in our model are tuned, thereby verifying the robustness of our theory. Sec.~\ref{subsec:ExperimentalComparison} is devoted to the comparison of the theoretical profiles with the experimental analysis of the 
interfacial structure. In addition, we calculate the line tension of the free edge and the $\pi$-wall as a function of temperature and discuss the relative thermodynamic stability of the $\pi$-walls with respect to the monolayer edge. Concluding remarks are offered in the final section of the paper.

\section{Theory}
\label{sec:Theory}

\subsection{Structures of the $\pi$-wall and the monolayer edge}
\label{subsec:Structure}

\begin{figure}[h]
\centering
\includegraphics[scale=0.379]{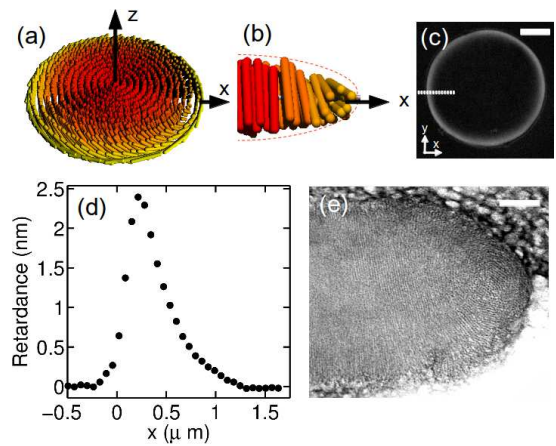}
\caption{{\bf Structure of the membrane edge.} (a) The schematics of the rod orientation in a membrane and (b) at its edge~\cite{gibaud}. (c) Retardance image of a membrane (top view)~\cite{Barry}. Black region corresponds to no retardance, whereas the bright ring at the edge is birefringent. (d) Radial retardance profile as a function of distance from the membrane edge (on the white dotted line shown in (c))~\cite{Barry}. (e) Electron micrograph of the cross-section of the membrane edge~\cite{gibaud}. It visualizes both the curved shape and the Sm-$A^\ast$-to-$Ch$ transition towards the edge. Scale bars, 5 $\mu$m (c); 0.2 $\mu$m (e). }
\label{fig:monolayer}
\end{figure}

Disk-shaped assemblages of $fd$-viruses typically reach mesoscopic sizes in diameter~\cite{Barry2}. Because of the tendency of monodisperse hard rods to build a one-rod-length thick flat monolayer, Sm-$A^\ast$ membranes do not support molecular twist and bend deformations and thus expel them to the edge (Figs.~\ref{fig:monolayer}a,~\ref{fig:monolayer}b)~\cite{DeGennes1}. This constraint is mediated by the depletion interactions between the rods and the depletant polymer. By contrast, in a conventional smectic-$A$ material, the exclusion of twist and bend deformations in the ground state is a natural consequence of constant distance everywhere between two adjacent layers. In general, bend deformations of the molecules localized at the edge are negligible as the rod length is much smaller than the monolayer radius. When $\lambda_b\ll r$ and $\bar{t}\ll r$, the contribution $\gamma_b$ of molecular bend to the line tension scales as $\gamma_b\sim K_3 \bar{t}\lambda_b /r^2\rightarrow 0$, where $K_3$ is the bend 
elastic constant, $\bar{t}$ is the average membrane thickness where the bend distortions occur, $\lambda_b$ is the penetration depth of the bend deformations at the edge, and $r$ is the radius of the monolayer. Since the rods are perfectly aligned in the bulk, the membrane interior appears black when viewed under 2D LC-Polscope, corresponding to no birefringence (Fig.~\ref{fig:monolayer}c)~\cite{Barry, gibaud}. Towards the edge, however, optical anisotropy arising from the twist deformations results in a bright, birefringent region characterized by a retardance, R, which is given by (Fig.~\ref{fig:monolayer}d)

\begin{equation}
\label{eq:retardance}
R=t\Delta n_{sat}cS\sin^2{\theta}.
\end{equation}
Here $t$ is the local membrane thickness. $n_{Sat}=(3.8\pm0.3)\times10^{-5}$mL/mg is the specific birefringence of a fully aligned bulk sample at unit concentration~\cite{purdy}. The nematic order parameter $S$ was measured to be 0.95 at a rod concentration $c$ of 100 mg/mL in the bulk monolayer~\cite{purdy}. Therefore we assume that $S\sim1$, namely there is perfect orientational order~\cite{Barry, Barry2}. $\theta$ is the tilt angle of the molecules with respect to the layer normal. 

To quantify the retardance, a simple model focusing on the spatial variation of the nematic director was proposed~\cite{Barry}. This model neglects the variation of the thickness and smectic order across the monolayer. For chiral membranes, the theory accurately reproduces the experimental retardance profile. However, as evidenced by their retardance map, achiral monolayers exhibit twist at the edge as well, spontaneously breaking the chiral symmetry at the membrane edge~\cite{gibaud}. Furthermore, electron micrographs of the membrane edge cross-section reveal a curved thickness profile and melting of the smectic order into a narrow $Ch$ band at the edge. In Fig.~\ref{fig:monolayer}e, the rods are seen to be perfectly aligned in the $z$-direction in the bulk of the monolayer. Towards the edge, the rods tilt rapidly such that they point perpendicular to the image plane, forming a $Ch$ region with a hemi-toroidal curved shape at the edge. 

\begin{figure}[h]
\centering
\includegraphics[scale=0.4]{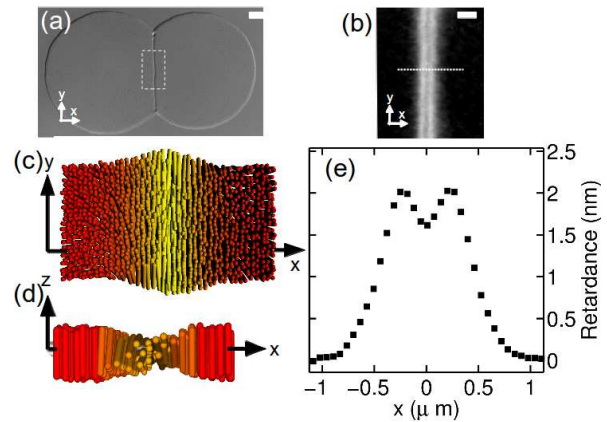}
\caption{ {\bf Structure of a $\pi$-wall~\cite{Zakhary}.} (a) Two coalesced monolayers generate a $\pi$-wall in between. (b) Retardance image of a $\pi$-wall, showing a narrow black band at the center surrounded by a bright region. Schematics of the $\pi$-wall; (c) top view, (d) side view. (e) Retardance profile versus distance from the $\pi$-wall (on the white dotted line shown in (b)). The narrow black region in (b) corresponds to a local minimum at the center. Scale bars, 3 $\mu$m (a); 1 $\mu$m (b). }
\label{fig:linedefect}
\end{figure}

This $Ch$ band constitutes the core of the $\pi$-wall as well, since a $\pi$-wall is formed from two free edges of same handedness (Fig.~\ref{fig:linedefect}a). Therefore, we model the $\pi$-wall by a $Ch$ region sandwiched between two semi-infinite Sm-$A^\ast$ monolayers (Figs.~\ref{fig:linedefect}c,~\ref{fig:linedefect}d). $\pi$-walls are linear defects, thereby their liquid crystalline order does not possess any bend deformations. The in-plane curvature of a $\pi$-wall is always vanishingly small compared to the edge curvature of a monolayer membrane~\cite{Zakhary}. Thus, the contribution of the bending term to the line tension of the $\pi$-walls is negligible, thereby not affecting their stability. As for their cross-sectional shape, the effect of the thickness variation is anticipated to be more prominent at the $\pi$-walls: an intensity profile having a local minimum located between two peaks is extracted from retardance measurements (Figs.~\ref{fig:linedefect}b,~\ref{fig:linedefect}e). We hypothesize 
that this local minimum indicates a significant variation of the membrane thickness across the $\pi$-wall.	

The presence of the depleting agent results in an effective surface tension over the membrane surface. When a single molecule protrudes from the monolayer, it reduces the accessible volume of the surrounding depletant polymer. Both in experiments and simulations, it was found that protrusion fluctuations are significant in Sm-$A^\ast$ membranes, and stabilize this phase against the configuration where single disks stack up on top of each other and form a layered Sm-$A^\ast$ structure~\cite{Barry2, Yang, Yang2}. In addition, spontaneous twist at the edge of achiral monolayers is understood as an outcome of the surface tension: compared to an untilted edge configuration, the monolayer may reduce the rod-polymer interfacial area despite resulting in an elastic energy associated with twist deformations~\cite{gibaud}. This configuration lowers the total surface energy of the membrane, especially if the surface tension continuosly varies throughout the monolayer and becomes bigger at the edge than in the bulk, due 
to the changing rod orientation. This scenario will be discussed below. Hence, a curved edge shape is favored when surface tension becomes dominant in determining the structure of the monolayer edge. In previous theories, the thickness variation, as well as the $Ch$-band formation, were not taken into account, resulting in a line tension which is low by an order of magnitude in comparison to experimental results~\cite{Pelcovits, Kaplan}.

\subsection{The unified free energy}
\label{subsec:FreeEnergy}

These recent experimental findings are beyond the scope of the simple model proposed in Ref.~\cite{Barry}. Thus, in addition to the spatial change of the molecular tilt $\theta$, a detailed theory should account for the local thickness variation $t$ and the crossover from Sm-$A^\ast$-to-$Ch$ behavior $\Psi$ across the monolayer and the $\pi$-wall, in order to correctly predict their structure as well as the magnitude and the behavior of the line tension. 

For a bulk Sm-$A^\ast$ sample, $\Psi$ is defined as the thermodynamic average of the amplitude of a periodic mass-density wave, which preserves the discrete translational symmetry in the direction locally parallel to the normal vector of the layers. A colloidal monolayer composed of rod-like particles lacks this symmetry. However, towards the edge monolayers exhibit a clear transition from an ordered phase where the molecules are aligned along the layer normal, to a $Ch$ region in which the molecules satisfy the chiral interactions. Therefore we model the monolayer as a binary fluid, in which the Sm-$A^\ast$ and $Ch$ liquid-crystalline regions coexist, separated by a diffuse interface with a coherence length $\xi_{||}$. In our model $\Psi$ is proportional to the local Sm-$A^\ast$ concentration, hence it is a non-conserved order parameter field.  $\Psi=\Psi_0$ is a perfect Sm-$A^\ast$ sample, whereas at $\Psi=0$  rods form $Ch$. On the contrary, the thickness $t$ is a conserved field, as the volume of the 
monolayer, which is the integral of the thickness over the monolayer domain, is conserved. In summary we have two non-conserved fields $\theta$ and $\Psi$, and a conserved field $t$. Free energy functionals coupling conserved and non-conserved fields is common in physical modeling, such as the growth of a stable phase in a supersaturated phase~\cite{Elder}. Thus, our approach can be classified as an equlibrium counterpart of model-C dynamics, which is the class of dynamic models containing both conserved and non-conserved fields~\cite{Halperin}.

The free energy per unit length of the Sm-$A^\ast$ flat layer in $(x, y)$-plane is given by
\begin{equation}
\begin{split}
\label{eq:theory1}
F=&\int dx\hskip4pt |t|\left[f_{Ch-A}+\frac{C_t}{2}\Psi^2(t^2-(t_0 \cos\theta)^2)^2-\nu\right]\\&+\oint ds\hskip4pt\left(\sigma_{||}(\mathbf{\hat{N}}\cdot\mathbf{n})^2+\sigma_{\perp}(\mathbf{\hat{N}}\times\mathbf{n})^2+k\kappa_c^2\right)\,.
%+\nu\left(t_0-t\right	)
\end{split}
\end{equation}

\begin{figure}[h]
\centering
\includegraphics[scale=0.4]{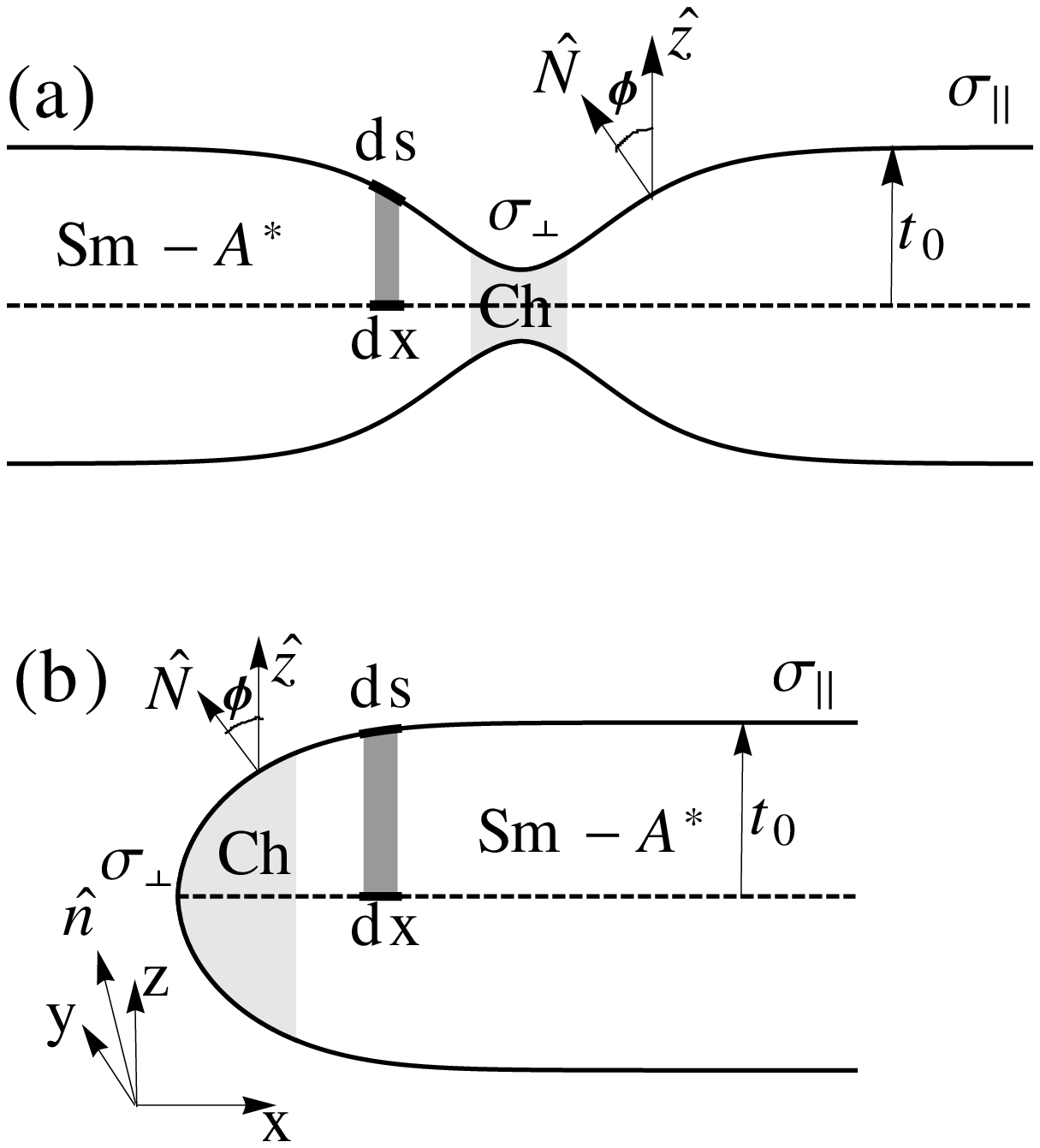}
\includegraphics[scale=0.4]{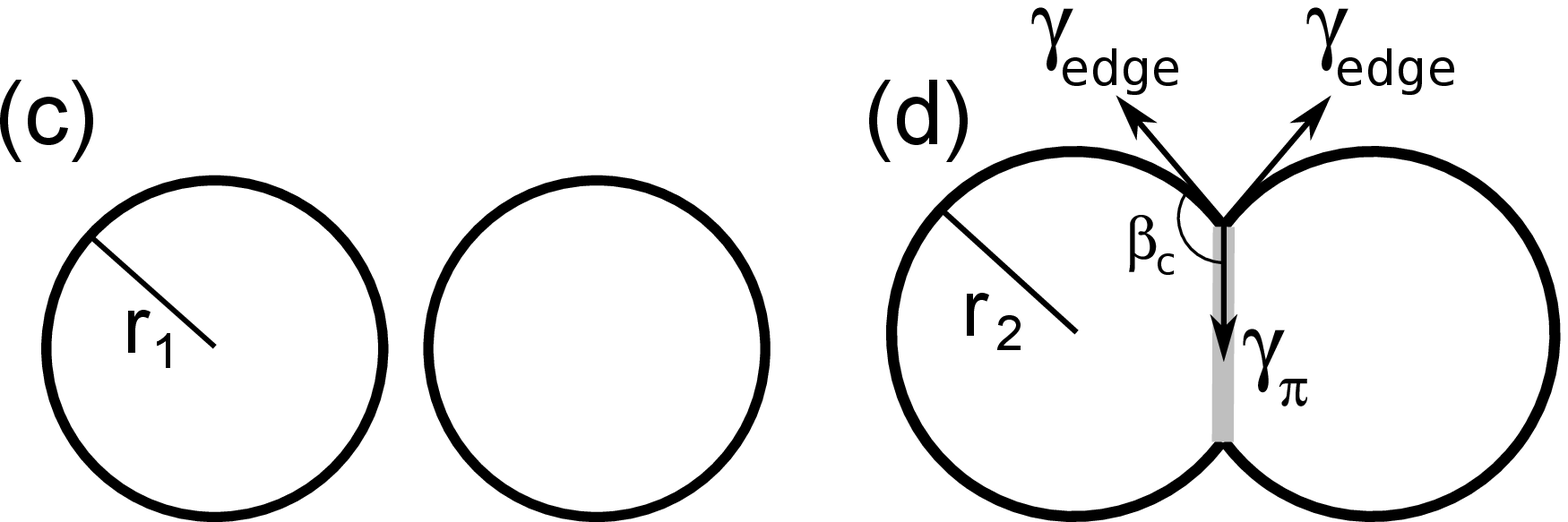}
\caption{{\bf Schematics of the $\pi$-wall and the edge.} Side views of (a) the $\pi$-wall and (b) the edge. Dashed lines denote the mid-plane. Light-shaded area indicates the $Ch$ region. Dark-shaded area indicates the projection of ds, which is dx. The surface normal $\mathbf{\hat{N}}$ is in the $(x, z)$-plane. $\phi$ is the angle between the z-axis and the surface normal $\mathbf{\hat{N}}$. The nematic director $\mathbf{n}$ is always in the yz-plane. Schematics of (c) the two monolayers of radii $r_1$ before coalescence and (d) the coalesced final structure with a $\pi$-wall (gray line). Each arc has the radius $r_2$. The line tensions $\gamma_{edge}$ and $\gamma_{\pi}$, equivalent to the applied forces at the anchoring point, should be balanced in equilibrium. $\beta_c$ is the contact angle. The relation $r_2\geq r_1$ always holds. }
\label{fig:sketch}
\end{figure}

\noindent The volume terms given in square brackets are multiplied by the local monolayer thickness $t(x)$. The only symmetry related with $t$ is up-down symmetry across the monolayer (dashed lines in Figs.~\ref{fig:sketch}(a) and (b)).  Since we calculate the profiles above the symmetry axis, we replace the volume element $|t| dx$ by $t dx>0$ in the following, without loss of generality. $f_{Ch-A}$ in Eq.~\eqref{eq:theory1} is written as
\begin{equation}
\begin{split}
\label{eq:theory2}
f_{Ch-A}=&-\frac{r}{2}\Psi^2+\frac{u}{4}\Psi^4+\frac{C_1}{2}\left(\nabla\Psi\right)^2+\frac{C_2}{2}(\Psi \sin\theta)^2\\&+\frac{K_2}{2} (\mathbf n\cdot\nabla\times\mathbf n-q_0)^2\,.
\end{split}
\end{equation}
Eq.~\eqref{eq:theory2} resembles the de Gennes free energy for the $Ch$-Sm-$A^\ast$ transition~\cite{DeGennes1, DeGennes2, Lubensky}, except for the fourth term, which accounts for the \textit{big} distortions of the nematic director in a flat smectic monolayer~\cite{Barry, Kaplan}. Furthermore, $\Psi$ is a real scalar field. The first three terms in Eq.~\eqref{eq:theory2} describe the transition from one rod-length thick monolayer of perfectly aligned $fd$ ($\Psi=\Psi_0=\sqrt{\frac{r}{u}}$) to the $Ch$ region forming at the monolayer edge and around the $\pi$ wall ($\Psi=0$). The energy cost due to big distortions of the nematic director $\mathbf n$ and arising from $t\neq\cos{\theta}$ towards the $\pi$ wall and the edge is included in the second term of Eq.~\eqref{eq:theory1} and in the fourth term of Eq.~\eqref{eq:theory2} without loss of generality, ensuring rotational invariance. The fourth term in Eq.~\eqref{eq:theory2}, which penalizes the tilting of $\mathbf{n}$ distorting perfect layer formation, 
depends nonlinearly on the tilt angle. This term has no counterpart in superconductors. The last term of Eq.~\eqref{eq:theory2} represents the twist deformations, where $K_2$ is the twist elastic modulus. It is the only contribution to $f_{Ch-A}$ from the Frank free energy density, as the nematic director $\mathbf{n}\equiv\left(0, \sin\theta(x), \cos\theta(x)\right)$ rotates by an angle $\theta$ about the $x$-axis within a characteristic length scale $\lambda_t\equiv\sqrt{\frac{K_2}{C_2\Psi_0^2}}$, the twist penetration depth~\cite{Barry, DeGennes1}. Furthermore, the second term in Eq.~\eqref{eq:theory1} dictates that $t=t_0$ (see Figs.~\ref{fig:sketch}a,~\ref{fig:sketch}b) in the region of fully aligned $fd$ viruses ($\Psi=\Psi_0$), and $t$ is decoupled from the orientation of $\mathbf n$ at the $Ch$ band ($\Psi=0$). The last volume term denoted by the Lagrange multiplier $\nu$ is due to the volume constraint of the membrane, arising from the rigid nature of the rod-like $fd$ viruses. That is, the overall 
monolayer volume stays constant, being independent of the local molecular orientation. 

The effect of the depletant polymer is represented by surface tension terms given in the second line of Eq.~\eqref{eq:theory1}. The coefficients in the fourth and fifth terms of Eq.~\eqref{eq:theory1} are the bulk surface tension $\sigma_{||}$ when $\mathbf{n}\parallel\mathbf{\hat{N}}$, and the $\pi$-wall or edge surface tension $\sigma_\perp$ when $\mathbf{n}\perp\mathbf{\hat{N}}$, $\mathbf{\hat{N}}\equiv(-\sin\phi(x), 0, \cos\phi(x))$ being the local unit normal of the membrane curved surface. $\phi=\tan^{-1}\frac{dt}{dx}$ is the angle between $\mathbf{\hat{N}}$ and $z$-axis. $ds$ is the infinitesimal arc length of the curved surface, and its projection onto the \emph{x}-axis is $dx$ (Figs.~\ref{fig:sketch}a,~\ref{fig:sketch}b). In the presence of anisotropy ($\sigma_{||}\neq\sigma_\perp$), the local surface tension changes continuously due to the local tilt of $\mathbf n$ between the two regimes. The last term in Eq.~\eqref{eq:theory1} is the curvature free energy cost of the surface~\cite{Helfrich1}, 
where $k$ is the associated curvature modulus. This term becomes particularly important in calculating curved edge profiles in agreement with Fig.~\ref{fig:edge}(e). For the $\pi$-walls, on the other hand, when ignoring the curvature energy cost, Eq.~\eqref{eq:theory1} still produces accurate retardance profiles as a function of the distance from the interface and surface energies as a function of chirality. Instead of using $t(x)$ and its derivatives, it is preferable to parametrize the curvature by the derivative of $\phi$ with respect to the infinitesimal arc length, that is $\kappa_c\equiv\frac{d\phi}{ds}$, where $ds=\frac{dx}{\cos\phi}$~\cite{Helfrich1}.

The necessity to include three scalar fields, namely the tilt angle $\theta$, order parameter $\Psi$, and thickness $t$, is justified as follows: In order to determine the width of the cholesteric $d_{Ch}$ (shown in Figs.~\ref{fig:edge} and~\ref{fig:piwall}), which is experimentally seen (Fig.~\ref{fig:monolayer}(e)), we need to calculate the fields $t$ and $\Psi$ simultaneously. When these two fields are obtained as a function of $x$, then the integration of  $\int^L_0 t\Psi/\Psi_0 dx$ over the monolayer domain $L$ gives the volume of the Sm-$A^\ast$ phase. The remaining volume is occupied by the $C$h phase, and its distribution underneath the curved thickness profile $t$ yields $d_{Ch}$. The determination of the cholesteric width is only possible when both $t$ and $\Psi$ are taken into account. Furthermore, the molecular tilt configurations minimizing the Sm-$A^\ast$ and the $Ch$ free energy are different. In a $Ch$ there is a cholesteric axis along which $\nabla\times\mathbf{n}=q_0$, 
whereas  $\nabla\times\mathbf{n}=0$ in a perfect Sm-$A^\ast$. Additionally, the local monolayer thickness is given by the $t_0 \cos{\theta}$ in the Sm-$A^\ast$, whereas it is decoupled from $\theta$ in the $Ch$ regime. The crossover between the two regimes is maintained by $\Psi$.

Taking the monolayer domain $L\rightarrow\infty$, and substituting the expressions for $\mathbf{n}$, $\mathbf{\hat{N}}$, and $\kappa_c$, Eqs.~\eqref{eq:theory1} and~\eqref{eq:theory2} are rewritten as
\begin{equation}
\begin{split}
\label{eq:theory3}
F=&\int_0 ^\infty dx\hskip4pt \left[t f_{Ch-A}+\frac{C_t}{2}t \Psi^2(t^2-t_0^2\cos^2\theta)^2-\nu t\right.\\&\left.+\frac{\sigma_{\perp}}{\cos\phi}-\Delta\cos^2\theta\cos\phi+k\cos\phi\left(\frac{d\phi}{dx}\right)^2\right.\\&\left.+\mu\left(\tan{\phi}-\frac{dt}{dx}\right)\right]\,,
%+\nu\left(t_0-t\right)
\end{split}
\end{equation}
where $\Delta\equiv\sigma_{\perp}-\sigma_{||}$, and

\begin{equation}
\begin{split}
\label{eq:theory4}
f_{Ch-A}=&-\frac{r}{2}\Psi^2+\frac{u}{4}\Psi^4+\frac{C_1}{2}\left(\frac{d\Psi}{dx}\right)^2+\frac{C_2}{2}\Psi^2\sin^2\theta\\&+\frac{K_2}{2} \left(\frac{d\theta}{dx}-q_0\right)^2\,.
\end{split}
\end{equation}
In Eq.~\eqref{eq:theory3}, due to the parametrization of $\mathbf{\hat{N}}$ and $\kappa_c$ in terms of $\phi$, there is an extra constraint of $\tan{\phi}=\frac{dt}{dx}$ multiplied by the Lagrange multiplier $\mu=\mu(x)$.

For one independent and several dependent variables, the Euler-Lagrange (EL) equations which minimize Eq.~\eqref{eq:theory3} are given by~\cite{Arfken}
\begin{equation}
\label{eq:EL}
\frac{\partial f}{\partial g_i}=\frac{d}{dx}\frac{\partial f}{\partial g'_i}
\end{equation}
where $g_i\equiv\{\theta, \Psi, t, \phi\}$, $g'_i\equiv\{\theta', \Psi', t', \phi'\}$, and $f=f(g_i, g'_i)$ is the free energy density given by $F=\int dx f(g_i, g'_i)$. Primes denote derivatives $\frac{dg}{dx}$. With the constraint given in the last term of Eq.~\eqref{eq:theory3},  Eq.~\eqref{eq:EL} contains eight coupled nonlinear first-order differential equations that are solved subject to eight boundary conditions. These eight equations are calculated as
\begin{equation}
\label{eq:EL2a}
u_1\equiv\frac{\partial f}{\partial \theta'}=K_2 t \left(\theta'-q_0\right)\,,
\end{equation}
\begin{equation}
\label{eq:EL2b}
u_2\equiv\frac{\partial f}{\partial \Psi'}=C_1 t \Psi'\,,
\end{equation}
\begin{equation}
\label{eq:EL2c}
t'=\tan\phi\,, 
\end{equation}
\begin{equation}
\label{eq:EL2d}
\kappa_c=cos\phi \phi'\,,
\end{equation}
\begin{equation}
\label{eq:EL2e}
u_1'=\sin{2\theta}\left[\Delta\cos\phi+t\Psi^2\left(\frac{C_2}{2}-C_t t_0^4 \cos^2\theta+C_t t_0^2 t^2\right)\right]\,, 
\end{equation}
\begin{equation}
\label{eq:EL2f}
u_2'=t\Psi\left[-r+C_2\sin^2\theta+C_t\left(t^2-t^2_0\cos^2\theta\right)^2+u\Psi^2\right]\,, 
\end{equation}
\begin{equation}
\label{eq:EL2g}
\begin{split}
 \mu'=&\nu-\frac{\Psi^2}{2}\left[-r+C_2\sin^2\theta+C_t t_0^4\cos^4\theta\right.\\&\left.-6 C_t t_0^2 t^2 \cos^2\theta+5 C_t t^4\right]-\frac{u}{4}\Psi^4-\frac{u_1^2}{2 K_2t^2}-\frac{u_2^2}{2 C_1t^2}\,,
\end{split}
\end{equation}
\begin{equation}
\label{eq:EL2h}
\begin{split}
2k\cos^2\phi\kappa_c'=&\Delta\cos^2\theta\sin\theta\cos^2\phi+\sigma_{\perp}\sin\phi\\&+\mu-k\sin\phi\kappa_c^2\,.
\end{split}
\end{equation}

The boundary conditions in the bulk are specified as follows: because there is parallel alignment of rods in the interior of the monolayer, no director tilt occurs ($\theta_0=0$ or $\theta_0=\pi$), there is perfect smectic order ($\Psi=\Psi_0$), and $t=t_0$ (see Figs.~\ref{fig:sketch}a and~\ref{fig:sketch}b). The unit layer normal $\mathbf{\hat{N}}$ is parallel to the $z$-axis, leading to $\phi_0=0$. Furthermore, all derivatives must vanish in the bulk in the ground state of the monolayer.

Using the boundary conditions in the bulk, we determine the Lagrange multiplier of the volume constraint from Eq.~\eqref{eq:EL2g} as

\begin{equation}
\label{eq:multiplier}
\nu=\frac{1}{2}\left(-\frac{r^2}{2u}+K_2q_0^2\right)\,.
\end{equation}
The physical meaning of $\nu$ can be understood by the following argument: the volume difference between a configuration of uniformly aligned viruses parallel to the $z$-axis and the configuration shown in Fig.~\ref{fig:monolayer}c contributes to the bulk of the monolayer by the associated energy density $\nu$, as the volume of the monolayer is constrained to stay constant in Eq.~\eqref{eq:theory1}. The first term in Eq.~\eqref{eq:multiplier} is the energy gain due to the Sm-$A^\ast$ order, and the second term is the energy cost of the chiral rods avoiding twist deformations. $\nu$ vanishes at $q_{0 C}=\Psi_0^2 \sqrt{\frac{u}{2 K_2}}$, where $q_{0 C}$ is defined as the critical chirality.  

For the $\pi$-wall and the monolayer edge structure, the boundary conditions are determined separately. The thickness of a $\pi$-wall saturates at a thickness $t_\pi$ at the center, where $\mathbf{\hat{N}}$ is parallel to the $z$-axis, hence $\phi_\pi=0$. Since $\pi$-walls are modeled by a $Ch$ region sandwiched between two Sm-$A^\ast$ monolayers, $\Psi_{\pi}=0$ and $\theta_\pi=\frac{\pi}{2}$ by symmetry. As for the boundary conditions at the edge, from the electron micrograph (Fig.~\ref{fig:monolayer}e), it is clear that $\phi_{edge}=\frac{\pi}{2}$, where $t_{edge}=0$. Additionally, at the edge the viruses stay parallel to the $y$-axis along the membrane periphery, that is $\theta_{edge}=\frac{\pi}{2}$. We assume the rods forming a perfect $Ch$ at the monolayer edge, thus $\Psi_{edge}=0$. 

The membrane edge and the $\pi$-wall show significant thermal fluctuations on the monolayer plane. The line tension $\gamma_{eff}$, which is the free energy cost associated with the formation of these interfaces, can be extracted from the analysis at the long wavelength limit of the measured fluctuation spectra~\cite{gibaud, Zakhary}. The magnitude of $\gamma_{eff}$ is controlled by the chirality of the constituent rods, since the twist deformations expelled to the interface reduce the energy of rods arising from chiral interactions. The fluctuation amplitude is inversely proportional to $\gamma_{eff}$. The bigger the depletant concentration, the lower the fluctuation amplitude becomes at long wavelengths, leading to an increase in $\gamma_{eff}$~\cite{Helfrich2}. Therefore it was proposed that the line tension is given by
\begin{equation}
\gamma_{eff}=\gamma_{bare}-\gamma_{chiral}\,,
\label{eq:gammaeff}
\end{equation}
where $\gamma_{bare}$ depends on the depletant concentration, and $\gamma_{chiral}$ is the chiral contribution to the line tension~\cite{gibaud}. Theoretically, $\gamma_{eff}$ can be calculated by~\cite{Safran} 
\begin{equation}
\gamma_{eff}=F-L t_0 f_0\,,
\end{equation}
where F is given by Eq.~\eqref{eq:theory1}, and the bulk free energy density of a large disk is written as $f_0=\nu+\sigma_{||}/t_0$, since in Eq.~\eqref{eq:theory1} the contributions of twist deformations to $f_0$ are negligible~\cite{Pelcovits, Kaplan}. Rods present inside the bulk create a surface energy associated with $\sigma_{||}$, which results in the second term of $f_0$.

\subsection{Thermodynamic stability of the free edge and the $\pi$-wall}
\label{subsec:coalescence}

Since the formation of $\pi$-walls can simply be induced by the coalescence of two membranes~\cite{Zakhary}, we investigate the thermodynamic stability of two disconnected membranes with respect to two coalesced ones with a $\pi$-wall inbetween. The configurations of two monolayers before and after coalescence are shown in Figs.~\ref{fig:sketch}(c) and~\ref{fig:sketch}(d). The coalesced form adopts a $\pi$-wall in between (gray line). The magnitudes of the forces applied to the anchor points by the $\pi$-wall and the monolayer edge are equal to the $\gamma_{\pi}$ and $\gamma_{edge}$, the line tensions associated with the $\pi$-wall and the edge, respectively. In equilibrium these forces are balanced, giving rise to a certain contact angle $\beta_c$, which satisfies the relation
\begin{equation}
\gamma_{\pi}=-2\gamma_{edge}\cos\beta_c\,.
\label{eq:contact}
\end{equation}
Ignoring the detailed structure of the $\pi$-wall and the monolayer edge, we further assume that the total area of two monolayers is conserved. Namely,  $A_{1st}=A_{2nd}$, where the $A_{1st}$ and $A_{2nd}$, the total areas of the first and second configurations, respectively, are given by 
\begin{equation}
A_{1st}=2\pi r_1^2\,,\quad\text{and}\quad A_{2nd}=(2\beta_c-\sin{2\beta_c})r_2^2\,.
\label{eq:areas}
\end{equation}
To determine the stability of $\pi$-walls with respect to the membrane edge, we compare the total line energies of the first and second configurations, $F_{1st}$ and $F_{2nd}$, which are written as
\begin{gather}
F_{1st}=4\pi r_1\gamma_{edge}\,,
\\
F_{2nd}=4\beta_c r_2\gamma_{edge}+2r_2\sin{\beta_c}\gamma_{\pi}. 
\label{eq:simpleF}
\end{gather}
When $\Delta F\equiv F_{1st}-F_{2nd}>0$ under the constraints given by Eqs.~\eqref{eq:contact} and~\eqref{eq:areas}, $\pi$-walls should be favored. On the other hand, if $\Delta F\leq 0$, the $\pi$-wall length goes to zero as the two coalesced disks will separate into two distinct monolayers, namely back to the first configuration. At $\Delta F= 0$, $\beta_c$ becomes equal to $\pi$, implying a continuous transition between the two regimes. Hence, the region of $\pi$-wall stability is given as $\beta_c<\pi$. In this regime, the radius of a single arc in the second configuration, $r_2$, is always bigger than $r_1$, the radius of a single disk.

\section{Results}
\label{sec:Results}

\subsection{Methods}
\label{subsec:Methods}

We now explore the results by solving the EL Equations given in Eqs.~\eqref{eq:EL2a}--\eqref{eq:EL2h} and their respective boundary conditions for the $\pi$-wall and the monolayer edge. Since the EL Equations are coupled to each other and nonlinear, they are analytically not tractable. Instead, we use the relaxation method for boundary-value problems, where we replace the ordinary differential equations by a set of equivalent finite-difference equations on a grid of $M$ points. The origin of the grid is either the monolayer edge or the center of the $\pi$-wall, and the final boundary is taken to be large enough to ensure that all derivatives in Eqs.~\eqref{eq:EL2a}--\eqref{eq:EL2h} vanish, and all boundary conditions in the bulk of the Sm-$A^\ast$ membranes are satisfied. Starting from an initial guess, the algorithm iterates the solution by using Newton's method until the generated numerical values of the dependent variables converge to the true solution up to a relative error~\cite{NR}. In our 
implementation each of the two adjacent points on the grid are coupled, $M=800$, and the relative error is in the order of $10^{-8}$. The dependent variables to be simultaneously solved in our analysis are 
$\theta, \Psi, t,  \phi, \mu, \theta', \Psi',\text{and}\, \kappa_c$. 

When solving differential equations, it is suitable to work with dimensionless parameters. In our analysis, we use the assumption $t_0=\lambda_t$. The half-length of the $fd-wt$ viruses are $t_0=0.44\mu m$~\cite{Dogic}, and from previous theories $\lambda_t=0.48\mu m$~\cite{Barry}. In what follows, we will show that this assumption still holds.  Furthermore, for simplicity we assume that $r=C_2$. Hence, using the previous definitions of $\lambda_t\equiv\sqrt{\frac{K_2}{C_2\Psi_0^2}}$, $\Psi_0\equiv\sqrt{\frac{r}{u}}$, and measuring the distance from the monolayer edge and the $\pi$-wall in units of $\lambda_t$, we define the following dimensionless constants:
\begin{equation}
\nonumber 
\xi_{||}\equiv\sqrt{\frac{C_1}{r}}\,,\quad\kappa_2\equiv\frac{\lambda_t}{\xi_{||}}\,,\quad\alpha\equiv\sqrt{\frac{C_t t_0^4}{r}}\,,
\end{equation}
\begin{equation}
\label{eq:dimensionless}
\bar{\sigma}_{||}\equiv\frac{\sigma_{||}\lambda_t}{K_2}\,,\quad\bar{\sigma}_{\perp}\equiv\frac{\sigma_{\perp}\lambda_t}{K_2}\,,\quad\bar{\Delta}\equiv\frac{\Delta\lambda_t}{K_2}\,,\quad\bar{k}\equiv\frac{k}{K_2\lambda_t}\,,
\end{equation}
\begin{equation}
\label{eq:bulkdimensionless}
\quad\bar{\nu}\equiv\frac{\nu \lambda_t^2}{K_2}=-\frac{1}{4}+\frac{(q_0\lambda_t)^2}{2}\,.  
\end{equation}
Here $\xi_{||}$ is the coherence length, and  $\kappa_2$ the twist Ginzburg parameter~\cite{Lubensky}, which is analogous to the Ginzburg parameter $\kappa$ in superconductors~\cite{degennes3, sarma}. By definition, $\lambda_t$ is measured inside the Sm-$A^\ast$ region. That is, the width of the cholesteric band, which we define as $d_{Ch}$, does not contribute to $\lambda_t$ (see Figs.~\ref{fig:edge} and~\ref{fig:piwall}). $\alpha$ is the dimensionless coupling strength of the thickness $t$ with the projected height of a tilted rod, $t_0\cos\theta$, in the Sm-$A^\ast$ region. The remaining definitions denoted with bars are the corresponding dimensionless constants. $q_0\lambda_t$ in Eq.~\eqref{eq:bulkdimensionless} is the dimensionless chirality of the rods. Using the definitions given by the Eqs.~\eqref{eq:dimensionless},~\eqref{eq:bulkdimensionless} in Eqs.~\eqref{eq:theory3} and~\eqref{eq:theory4}, the dimensionless free energy per unit length $\bar{F}$ is obtained as $\bar{F}\equiv F/
K_2$. The dimensionless critical chirality is found as $q_{0c}\lambda_t=1/\sqrt{2}$~\cite{Lubensky}.

Given the definitions in Eqs.~\eqref{eq:dimensionless} and~\eqref{eq:bulkdimensionless}, the dimensionless forms of Eqs.~\eqref{eq:theory3} and~\eqref{eq:theory4} become
\begin{equation}
\begin{split}
\label{eq:theory3dimensionless}
F=&\int_0 ^\infty dx\hskip4pt \left[\frac{t}{2} f_{Ch-A}+\frac{\alpha}{2}t \Psi^2(t^2-\cos^2\theta)^2-\bar{\nu} t\right.\\&\left.+\frac{\bar{\sigma}_{\perp}}{\cos\phi}-\bar{\Delta}\cos^2\theta\cos\phi+\bar{k}\cos\phi\left(\frac{d\phi}{dx}\right)^2\right]\,,
%+\nu\left(t_0-t\right)
\end{split}
\end{equation}
where
\begin{equation}
\begin{split}
\label{eq:theory4dimensionless}
f_{Ch-A}=&-\Psi^2+\frac{\Psi^4}{2}+\frac{1}{\kappa_2}\left(\frac{d\Psi}{dx}\right)^2+\Psi^2\sin^2\theta\\&+\left(\frac{d\theta}{dx}-q_0\lambda_t\right)^2\,.
\end{split}
\end{equation}

The theoretical retardance profile is calculated by plugging the $t$ and $\theta$ profiles in Eq.~\eqref{eq:retardance}.  To successfully reproduce the retardance data using the model in Eq.~\eqref{eq:theory1}, the finite resolution of an object viewed with optical microscopy must be taken into account. The resolution is characterized by a Gaussian distribution function, and it is convolved with the theoretical retardance to compare the resulting profile with the experimental retardance data~\cite{Barry}. For best fit, the rod concentration $c$ in Eq.~\eqref{eq:retardance} and $\lambda_t$ are adjusted accordingly. Since the $\pi$-wall thickness $t_{\pi}$ is not determined by experiments, it is a fitting parameter as well. We extract $t_{\pi}$ from the local minimum of the retardance at the $\pi$-wall (Figs.~\ref{fig:linedefect}e, ~\ref{fig:unified}d).

To compare with the line tension measured in experiments, the $T$-dependence of $q_0$ allows us to calculate $\gamma_{eff}$, given in Eq.~\eqref{eq:gammaeff}, as a function of the temperature (Fig.~\ref{fig:energy}). At approximately $T_c=60$ $^oC$ wild-type $fd$ viruses become achiral~\cite{gibaud}. Hence $\gamma_{chiral}$ vanishes in Eq.~\eqref{eq:gammaeff}, leading to $\gamma_{eff}=\gamma_{bare}$. Beyond this critical temperature $T_c$ the experimental line tension shows no temperature dependence. The chirality of wild-type $fd$ viruses was converted into a temperature-dependent function, given by $q_0=a (T_c-T)^{1/2}$, where $a=0.056 ( ^oC \mu\text{m}^2)^{-1/2}$~\cite{gibaud}. Since self assembly of the rods into the monolayers is primarily governed by hard-core repulsions among the rods and the surrounding depletant, we assume that all other constants in Eqs.~\eqref{eq:theory1} and~\eqref{eq:theory2} are independent of temperature.

\subsection{The edge and $\pi$-wall profiles}
\label{subsec:ParameterTuning}

\begin{figure*}[ht]
\centering
\includegraphics[width=1\textwidth]{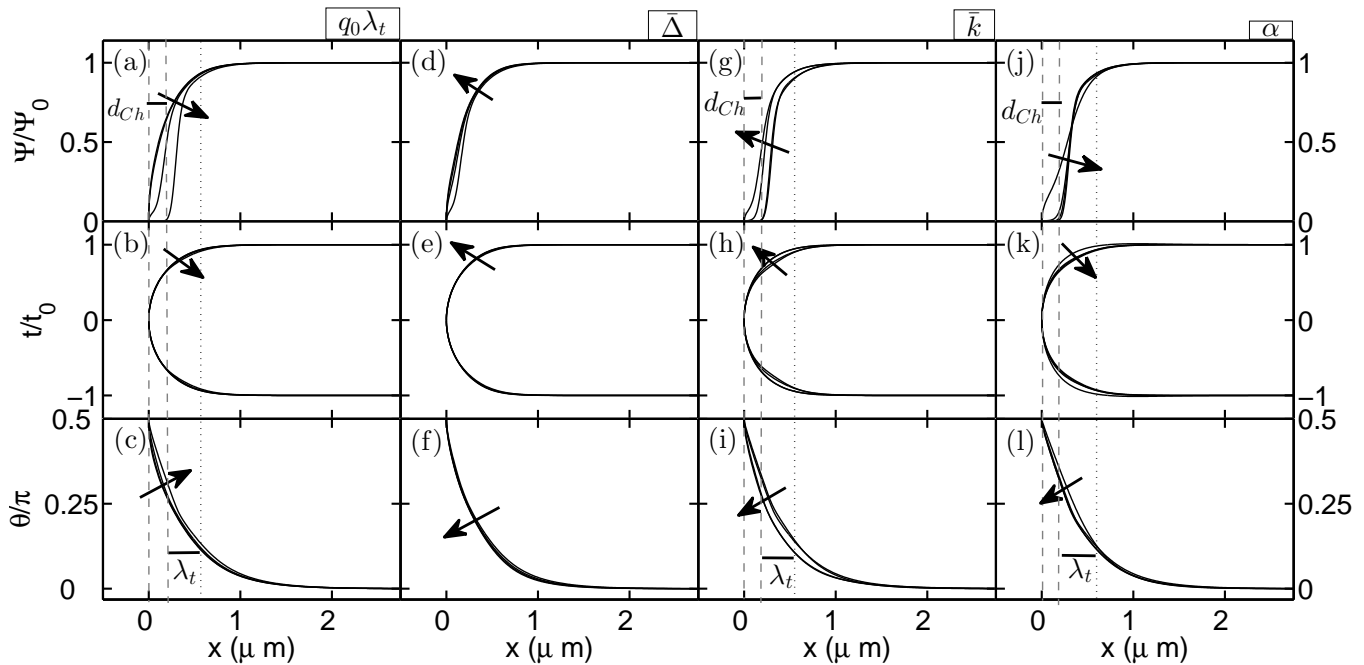}
\caption{{\bf From top to bottom, the profiles of the smectic order parameter $\Psi$, thickness $t$, and the tilt angle $\theta$, as a function of the distance from the edge.} $\Psi$ is normalized by $\Psi_0\equiv\sqrt{\frac{r}{u}}$, $t$ by $t_0=\lambda_t$, and $\theta$ by $\pi$ (vertical axes). The origin of the horizontal axes lie at at the monolayer edge, positive values of the distance are inside the monolayer. From left to right, each column displays the evolution of the profiles upon variation of the corresponding dimensionless variable; the chirality $q_0\lambda_t=\{0, 0.3, 0.6, 1/\sqrt{2}\}$, the surface tension anisotropy $\bar{\Delta}=\{0.1, 0.2, 0.3, 0.4\}$, the curvature modulus $\bar{k}=\{0.1, 0.2, 0.5, 1.0\}$, and the thickness coupling strength $\alpha=\{1, 5, 10, 15\}$, respectively (Eq.~\eqref{eq:dimensionless}), as denoted at the top. All other parameters are taken to be constant for each column. The arrows are oriented in the direction in which the profiles evolve as the corresponding dimensionless variable increases. In the first, third, and fourth columns, maximum width of the $Ch$ band $d_{Ch}$ and corresponding twist penetration depth $\lambda_t$ are shown. $\Psi$ exhibits $Ch$ region, $t$ qualitatively reproduces the curved shape in the electron micrograph (Fig.~\ref{fig:monolayer}b), and $\theta$ shows a decay characterizing the twist penetration at the edge~\cite{Barry}. Overall the profiles are fairly robust upon change of the variables in Eq.~\eqref{eq:theory1}.}
\label{fig:edge}
\end{figure*}

\begin{figure*}[ht]
\centering
\includegraphics[width=1\textwidth]{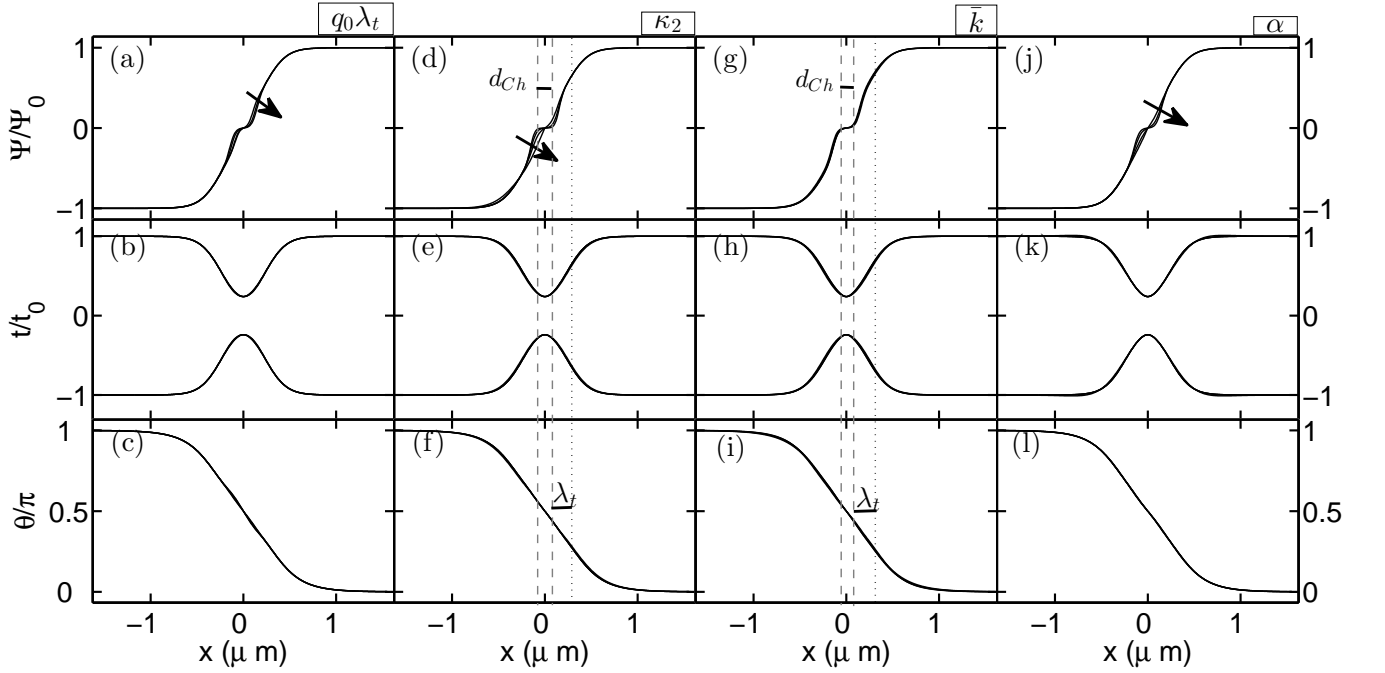}
\caption{{\bf From top to bottom, the profiles of the smectic order parameter $\Psi$, thickness $t$, and the tilt angle $\theta$, as a function of the distance from the $\pi$-wall. }
The origin of the horizontal axes lie at at the $\pi$-wall, positive and negative values of the distance are inside the monolayer in opposite directions. From left to right, each column displays the evolution of the profiles as a function of the dimensionless variables; the chirality $q_0\lambda_t=\{0.1, 0.3, 0.6, 1/\sqrt{2}\}$, the twist Ginzburg parameter $\kappa_2=\{2, 6, 10, 15\}$, the curvature modulus $\bar{k}=\{0.01, 0.02, 0.05, 0.1\}$, and the thickness coupling strength $\alpha=\{1, 5, 10, 15\}$, respectively (Eq.~\eqref{eq:dimensionless}), as denoted at the top. 
The arrows are oriented in the direction in which the profiles evolve as the corresponding dimensionless variable increases. In the second and third columns, maximum $d_{Ch}$ and corresponding $\lambda_t$ are shown. $\Psi$ exhibits the $Ch$ region across the $\pi$-wall, and $\theta$ shows a decay characterizing the twist penetration around the $\pi$-wall~\cite{Barry}. $t$ is significantly reduced from $t_0$ in the bulk to $t_0/4$ at the center of the $\pi$-wall.}
\label{fig:piwall}
\end{figure*}

Figs.~\ref{fig:edge}--\ref{fig:energy} display our main results. For convenience, in Figs.~\ref{fig:edge} and~\ref{fig:piwall} we illustrate the results in terms of dimensionless parameters to determine the relevant parameters which result in pronounced changes in the results. By minimizing Eqs.~\eqref{eq:theory3dimensionless} and~\eqref{eq:theory4dimensionless}, the smectic order parameter $\Psi$, the thickness $t$, and the tilt angle $\theta$ are calculated as a function of the following dimensionless parameters: the chirality $q_0\lambda_t$, the twist Ginzburg parameter $\kappa_2$, the bulk surface tension modulus $\bar{\sigma}_{||}$, the surface tension anisotropy $\bar{\Delta}$, the thickness coupling strength $\alpha$, and the curvature modulus $\bar{k}$. 

In Fig.~\ref{fig:edge} the evolution of  $\Psi$, $t$, and $\theta$, versus the distance from the edge, are shown as $q_0\lambda_t$, $\bar{\Delta}$, $\bar{k}$, and $\alpha$ are varied in successive columns, respectively. The set of these parameters is given in Fig.~\ref{fig:edge} caption, and all other parameters for a given column are taken to be constant. These constant parameters are; $q_0\lambda_t=q_{0c}\lambda_t=1/\sqrt{2}$, $\kappa_2=10$, $\bar{\sigma}_{||}=0.4$,  $\bar{\Delta}=0$, $\alpha=15$, and $\bar{k}=0.2$. To our knowledge, the critical chirality $q_{0c}\lambda_t$ does not have any thermodynamic significance for Sm-$A^\ast$ monolayers. However, $q_{0c}\lambda_t$ is fairly high compared to the range of $q_0\lambda_t$ observed in experiments which is approximately between $q_{0}\lambda_t\sim0-0.22$~\cite{gibaud}. Therefore a high value of $q_0\lambda_t$ allows us to  better examine the effects of the $Ch$ band formation to the edge structure. The effect of $\kappa_2$ and $\bar{\sigma}_{||}$ on the 
edge structure is found to be negligible. Thus, the evolution of the profiles with respect to $\kappa_2$ and $\bar{\sigma}_{||}$ are not shown.
From Eq.~\eqref{eq:dimensionless} the energy scale of the surface tension moduli is estimated as $K_2/\lambda_t\sim250 k_BT/\mu m^2$, given that $\lambda_t\sim0.5$ $\mu$m and $K_2=125\frac{k_BT}{\mu m}$~\cite{Dogic2, Barry}. This order of magnitude agrees well with the results presented in Ref.~\cite{Barry2}. Therefore we choose the bulk surface tension modulus to be $\bar{\sigma}_{||}=0.4$ (on the order of $10^2 \frac{k_BT}{\mu m^2}$) and assume the surface tension to be isotropic, that is $\bar{\Delta}=0$. $\alpha$ and $\bar{k}$ are not determined by experiments. We set $\alpha=15$ to work in the strong-coupling limit. Additionally, we find that $k\sim k_BT$ or $10k_BT$ results in $t$ profiles which qualitatively reproduce the shape of the edge as in Fig.~\ref{fig:monolayer}(d). This corresponds to a range of $\bar{k}\sim0.02-0.2$.

Figs.~\ref{fig:edge}(a)-(c) show the evolution of the profiles from the achiral limit ($q_0\lambda_t=0$) to $q\lambda_t=1/\sqrt{2}$. The reduction in the chiral energy density $-K_2 q_0 \mathbf{n}\cdot\left(\nabla\times\mathbf{n}\right)$, given by the cross-term in Eq.~\eqref{eq:theory2}, favors $d_{Ch}$ to become bigger when $q_0\lambda_t$ increases. This is also evidenced by the linear $\theta$ profiles. In the $Ch$ phase the last term in Eq.~\eqref{eq:theory3} is minimized when $\mathbf{n}\cdot\left(\nabla\times\mathbf{n}\right)=q_0$, leading to a linear $\theta$ profile, which corresponds to a uniform twist about a helical axis throughout the sample. The change in $t$ profiles is negligible, that is $q_0\lambda_t$ does not affect the overall shape of the monolayer edge.  The twist penetration depth is constant at $\lambda_t=0.38$ $\mu$m throughout the whole range of $q\lambda_t$. Likewise, in Ref.~\cite{gibaud} it is reported that the retardance profiles and $\lambda_t$ are independent of the molecular 
chirality $q_0$. 

In Figs.~\ref{fig:edge}(d)-(f), the effect of anisotropy in the surface tension ($\Delta>0$) is investigated. When anisotropy increases, $\theta$ profiles become more steep to suppress the effect of anisotropy, which is given by the fifth term in Eq.~\eqref{eq:theory1}. Accordingly $\Psi$ profiles show that $d_{Ch}$ decreases. Again, the shape of the monolayer edge is not affected, as evindenced by the $t$ profiles. The change in $\lambda_t$ is small; $\lambda_t=0.39$~$\mu$m for $\bar{\Delta}=0.1$ and $\lambda_t=0.42$ $\mu$m for $\bar{\Delta}=0.4$.

Figs.~\ref{fig:edge}(g)-(i) indicate that $\bar{k}$ has a profound effect on the shape and structure of the membrane edge. With increasing $\bar{k}$, the apices of $t$ profiles become flatter. Since we work in the strong coupling limit ($\alpha=15$), the change in $\bar{k}$ affects both $\theta$ and $\Psi$ profiles. $d_{Ch}$ becomes lower and $\theta$ profiles drift apart from the linear regime close to the edge. $\bar{k}$ significantly alters the twist penetration depth: $\lambda_t=0.34$ $\mu$m when $\bar{k}=0.1$ and $\lambda_t=0.47$ $\mu$m when $\bar{k}=1$.

The effect of the dimensionless coupling constant $\alpha$ is examined in Figs.~\ref{fig:edge}(j)-(l). As $\alpha$ increases, the apices of $t$ profiles become rounder.  $\theta$ profiles slightly change such that the twist penetration depth grows from $\lambda_t=0.34$ $\mu$m for $\alpha=1$ to $\lambda_t=0.38$ $\mu$m at $\alpha=15$. Additionally, $d_{Ch}$ becomes bigger when $\alpha$ increases. When $\alpha$ is small, there is a deviation between the projected height of the rod $t_0\cos{\theta}$ and $t$ in the Sm-$A^\ast$ region. On the other hand, bigger values of $\alpha$ result in good agreement between $t$ and $t_0\cos\theta$ outside $d_{Ch}$. Hence, our choice to work in the strong coupling limit is justified within the framework of our model.

In Fig.~\ref{fig:piwall} the evolution of  $\Psi$, $t$, and $\theta$, versus the distance from the edge, are shown as $q_0\lambda_t$, $\kappa_2$, $\bar{k}$, and $\alpha$ are varied in successive columns, respectively. Again, all other parameters for each column are taken to be constant. These constant parameters are the same as before, except that $\bar{k}=0.05$, that is, $k\sim3k_BT$. In contrast to the edge structure, $\pi$-walls are affected by the anisotropy $\bar{\Delta}$ negligibly. Therefore, the evolution of the profiles with respect to $\bar{\Delta}$ will not be discussed.  Overall the profiles are very robust upon change of the variables in Eq.~\eqref{eq:theory1}. For all columns except the column of $\bar{k}$, we extract $\lambda_t=0.21$ $\mu$m, which is fairly low compared to $\lambda_t$ of the edge, being independent of $q_0\lambda_t$. Namely the twist penetration depth is constant, as observed in experiments~\cite{gibaud}. 

The change in $q_0\lambda_t$ (Figs.~\ref{fig:piwall}(a)-(c)) and $\alpha$ (Figs.~\ref{fig:piwall}(j)-(l)) alter the profiles the same way as described above for the edge.  However, the effects of $\kappa_2$ and $\bar{k}$ on $\pi$-wall profiles are qualitatively different from the edge profiles. In Figs.~\ref{fig:piwall}(d)-(f), when $\kappa_2$ changes, the relaxation length scale $\xi_{||}$ of the smectic order affects $\Psi$ profiles. As a result, $d_{Ch}$ becomes bigger for higher $\kappa_2$, corresponding to smaller $\xi_{||}$, thereby rapid variation of $\Psi$. 
Moreover, Fig.~\ref{fig:piwall}(g)-(i) indicate that $\bar{k}$ has a negligible effect on the shape and structure of the $\pi$-wall. $\lambda_t=0.23$ $\mu$m (see Fig.~\ref{fig:piwall}(i)) for $\bar{k}=0.01$ and it reduces to 0.2 $\mu$m when $\bar{k}=0.1$.

\subsection{The retardance and the line tension}
\label{subsec:ExperimentalComparison}

\begin{figure}[h]
\centering
\includegraphics[width=1\columnwidth]{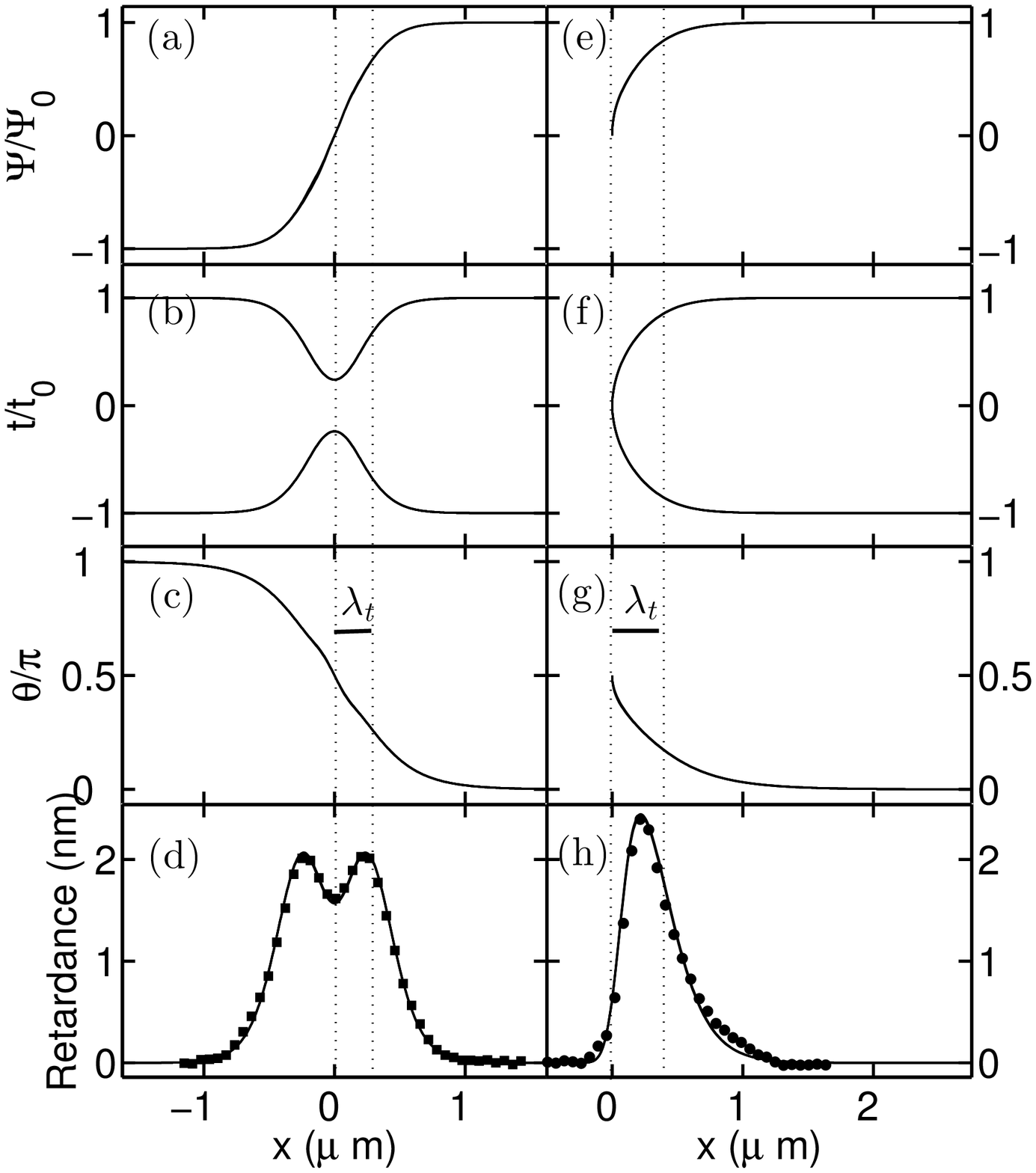}
\caption{ {\bf Experimental versus theoretical retardance.} From top to bottom, the profiles of the smectic order parameter $\Psi$, thickness $t$, the tilt angle $\theta$, and the retardance $R$, as a function of the distance from the $\pi$-wall (left column, (a)--(d)) and the edge (right column, (e)--(h)). The origins of the horizontal axes lie at the $\pi$-wall (left) and the edge (right). The dotted vertical lines indicate the region of twist penetration $\lambda_t$. The squares in (d) and dots in (h) are experimental retardance data, whereas full curves are calculated from Eqs.~\eqref{eq:EL2a}--\eqref{eq:EL2h}. Excellent agreement is obtained between experiment and theory, the theoretical retardance being robust upon the change of $q_0\lambda_t$ (or equivalently, temperature)~\cite{gibaud, Zakhary}. }
\label{fig:unified}
\end{figure}

Next we interpret the retardance profiles in Figs.~\ref{fig:unified}(d) and (h), calculated from the $t$ and $\theta$ profiles displayed in Figs.~\ref{fig:unified}(b)--(c) and (f)--(g). These profiles are obtained for the range of chirality $q_0\sim0-0.5$ $\mu$m$^{-1}$ (equivalently $T\sim0-60$ $^oC$) at the strong coupling limit ($\alpha=18$), corresponding to a depletant concentration of 35 mg/mL, as given in Tables~\ref{edgetable} and~\ref{piwalltable}. 
For both types of interfaces, Fig.~\ref{fig:unified} contains three sets of indistinguishable theoretical profiles (black full curves) corresponding to intermediate and high chiralities within this range, and the achiral limit. 
That is, the retardance profiles are essentially unchanged, revealing that the $\pi$-wall and the edge structures are independent of the rod chirality~\cite{gibaud}. The local minimum of the $\pi$-wall retardance is successfully reproduced by our model and allows us to extract the $\pi$-wall thickness as $t_{\pi}=t_0/4$ (Fig.~\ref{fig:unified}(d)). In the experimentally realized range of $q_0$, $\Psi$ profiles reveal that the width of the $Ch$ band goes to zero (Figs.~\ref{fig:unified}(a) and (e)). Furthermore, $\lambda_t$ is found to be bigger at the edge than at the $\pi$-wall (see Tables~\ref{edgetable},~\ref{piwalltable}). We have no simple physical reason for this difference. Revisiting our assumption that $t_0=\lambda_t$, for the edge and the $\pi$-walls, we obtain the membrane thickness of $t_0\simeq0.6$ $\mu$m and $t_0\simeq0.8$ $\mu$m, respectively. These results are in reasonable agreement with the expected membrane thickness of 0.88 $\mu$m, ignoring protrusion fluctuations of the rods.

\begin{table}[ht]
\centering
\begin{tabular}{c||c|c|c|c}
Dextran $\left(\frac{mg}{mL}\right)$ & $\sigma_{||} (\frac{k_BT}{\left(\mu\text{m}\right)^2})$ & $\sigma_{\perp} (\frac{k_BT}{\left(\mu\text{m}\right)^2})$ & $k (k_BT)$  & $\lambda_t (\mu\text{m})$  \\
\hline
 45 & 183 & 349 & 23.5 & 0.47 \\
  40 & 145 & 276 & 5.38 & 0.43 \\
  35 & 80 & 128 & 4.88 & 0.39
\end{tabular}
\caption{ For the edge, $\lambda_t$, surface tension and curvature moduli as a function of the depletant concentration. The twist elastic constant is kept fixed at $K_2$=125~$k_BT /\mu$m ~\cite{Dogic2}. The twist Ginzburg parameter is chosen as $\kappa_2=4$. Using Eq.~\eqref{eq:retardance}, the rod concentration is found to be $c=170$ mg/mL.}
\label{edgetable}
\end{table}

\begin{table}[ht]
\centering
\begin{tabular}{c||c|c|c|c}
Dextran $\left(\frac{mg}{mL}\right)$ & $\sigma_{||} (\frac{k_BT}{\left(\mu\text{m}\right)^2})$ & $\sigma_{\perp} (\frac{k_BT}{\left(\mu\text{m}\right)^2})$ & $k (k_BT)$  & $\lambda_t (\mu\text{m})$  \\
\hline
 45 & 184 & 350 & 2.13 & 0.34 \\
  40 & 161 & 282 & 1.94 & 0.31 \\
  35 & 112 & 179 & 1.75 & 0.28
\end{tabular}
\caption{ For the $\pi$-walls, $\lambda_t$, surface tension and curvature moduli as a function of depletant concentration ($K_2$=125~$k_BT\mu m$, $\kappa_2=4$, and $c=145$ mg/mL).}
\label{piwalltable}
\end{table}

\begin{figure}[h]
\centering
\includegraphics[width=1\columnwidth]{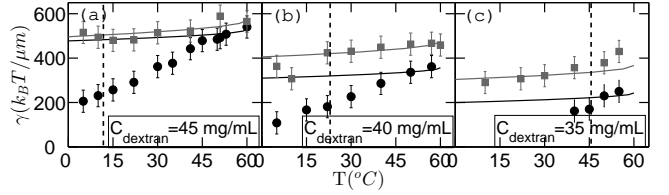}
\caption{{\bf The effective line tension $\gamma$ of both structures as a function of temperature $T$ and the depletant concentration $C_{dextran}$.} The  gray squares and black dots, with their error bars, are extracted from the experimental $\pi$-wall and edge fluctuation spectra, respectively~\cite{gibaud, Zakhary}. Likewise, the gray and black full curves are the theoretical $\gamma$ of the $\pi$-wall and the edge, respectively, calculated from Eq.~\eqref{eq:gammaeff}. The dashed lines correspond  $\beta_c=\pi$.} 
\label{fig:energy}
\end{figure}

Eq.~\eqref{eq:gammaeff} states that the chirality of the constituent rods controls the line tensions $\gamma_{edge}$ and $\gamma_{\pi}$. It follows that decreasing the temperature results in higher chirality of wild-type $fd$ virus, subsequently reducing $\gamma_{edge}$ and $\gamma_{\pi}$ (Fig.~\ref{fig:energy}), since the twist deformations favored by the edge and $\pi$-wall structures also satisfy the chiral interactions. This is realized in Figs.~\ref{fig:energy}(a)-(c). Furthermore, the depletant concentration $C_{dextran}$ affects the order of magnitude of the line tension. Consequently, when $C_{dextran}$ decreases from Figs.~\ref{fig:energy}(a) to (c), the overall line tension becomes smaller. 

In Fig.~\ref{fig:energy} the theoretical $\gamma_{edge}$ (black full curves) and $\gamma_{\pi}$ (gray full curves) are compared to the experiments (respectively, black dots and gray squares) as well. For the $\pi$-walls we obtain excellent agreement between theory and experiment over the entire range of temperature and depletant concentration. Chirality lowers $\gamma_{\pi}$ by up to 100 $k_BT/\mu$m. On the other hand, even though the low $q_0$ behavior and order of magnitude of theoretical $\gamma_{edge}$ agrees well with experiment, the slopes of the experimental and theoretical $\gamma_{edge}$ disagree. The origin of this discrepancy, as well as the slope difference between $\gamma_{edge}$ and $\gamma_{\pi}$ are not understood. Additionally, the theoretical line tension profiles of both structures are found to be close to parallel to each other. In contrast to the theory where the reduction is by up to 100 $k_BT/\mu$m, chirality reduces $\gamma_{edge}$ by e.g. up to 400 $k_BT/\mu$m when $C_{dextran}$=45 
mg/mL in experiments (Fig.~\ref{fig:energy}a).

The moduli $\sigma_{||}$, $\sigma_{\perp}$, and $k$ for the theoretical curves in Fig.~\ref{fig:energy} are given in Tables~\ref{edgetable} and~\ref{piwalltable} corresponding to the monolayer edge and the $\pi$-wall, respectively, as a function of the depletant concentration.  The magnitudes of $\sigma_{||}$ and $\sigma_{\perp}$ of both structures are found to be very close to each other, reflecting the unified nature of our model. Besides, their orders of magnitude match with experimental predictions~\cite{Barry2}. Whereas the curvature modulus $k$ is on the order of $k_BT$ for the $\pi$-wall, thermal fluctuations should not alter its shape, since the shape dependence on $k$ was found to be negligible (Figs~\ref{fig:piwall}(g)-(i)). For the edge, although $k$ affects the structure (see Figs.~\ref{fig:edge}(g)-(i)), the shape qualitatively agrees with the electron micrograph (Fig.~\ref{fig:edge}(e)) for the entire range of $k$.  Last, in order to fit the slope of $\gamma_{edge}$ to experiments, an extensive 
scan over the parameter space resulted in $\gamma_{edge}$ profiles of similar slopes, solely affecting their magnitude.  

Using Eq.~\eqref{eq:contact}, we extract the dependence of the contact angle $\beta_c$ on the experimental profiles of $\gamma_{edge}$ and $\gamma_{\pi}$. At $\beta_c=\pi$, $\Delta F$ vanishes (Eq.~\eqref{eq:simpleF}). Thus, at $T_c$ (the vertical dashed lines in Fig.~\ref{fig:energy}), which corresponds to $\beta_c=\pi$, $\pi$-walls should be separated continuously into two free edges and become unstable below $T_c$. Furthermore, when $T$ approaches $T_c$ from above, a gradual contraction of the $\pi$-wall length is expected. None of these predictions are confirmed in experiments, and $\pi$-walls are observed at temperatures as low as $T=5^oC$ (see Fig.~\ref{fig:energy}). Therefore, we conclude that $\pi$-walls survive, however become metastable at sufficiently low temperatures. In addition, we exclude the possibility of spontaneous formation of the $\pi$-walls at sufficiently high chirality, or equivalently at low temperatures. 

The anticipated dissociation of the coalesced configuration into two separate monolayers is replaced by the formation of alternating-bridge pore arrays (ABPAs), where the rods are aligned in the monolayer plane at the bridges~\cite{Zakhary}. The pores are occupied by the ambient suspension of depletant polymer. When ABPAs replace the $\pi$-walls, the amount of membrane-depletant interface considerably increases compared to a  $\pi$-wall. This interface has a curved two-dimensional shape accompanied with mean and Gaussian curvatures. When ABPAs form, they build a layered Sm-$A^\ast$ in the plane of the monolayer. Thus, in the presence of ABPAs, the structure is to be understood as an array of monolayer Sm-$A^\ast$--$Ch$--layered Sm-$A^\ast$--$Ch$--monolayer Sm-$A^\ast$, which has a counterpart neither in superconductors nor in conventional smectics. Modeling these structures is beyond the scope of our work, since our model is missing the relevant contributions due to the mean and Gaussian curvatures of the 
curved bridge-depletant interface.

\section{Conclusion}

The present theoretical study demonstrates that, in addition to the spatial change of the molecular tilt, the structure of the membrane edge and $\pi$-wall are strongly determined by the crossover between Sm-$A^\ast$ and $Ch$ regions, as well as the local thickness change in these structures. The membrane lowers the rod-depletant interfacial energy by a hemi-toroidal curved edge. On the other hand, the local thickness change of the $\pi$-wall occurs due to the global constraints imposed by its topology, resulting in a retardance drop at the $\pi$-wall. Our theory succesfully reproduces this unusual retardance behavior, confirming the hypothesized structure of the $\pi$-walls. Furthermore, the $\pi$-wall thickness, which is not measured in experiments, is determined by our model.

Our calculations indicate that the $\pi$-wall line tension is linear in the chirality of viruses. In contrast, line tension measurements of the edge revealed a quadratic dependence on the rod chirality~\cite{gibaud}, which is not well understood. A possible explanation is the effect of higher order chiral terms in the Frank elastic theory, as the variation of the molecular director field towards the edge is rapid and may be more significant than around the $\pi$-wall. 

In the superconductivity literature, the laminar model of alternating normal metal and superconducting regions was abondoned since the Abrikosov vortex lattice phase was found to be more favorable~\cite{goodman, sarma, degennes3}. Nevertheless, its analog in smectics is now realized in the coalescence of Sm-$A^\ast$ monolayers. Note that in the presence of big distortions Sm-$A^\ast$ monolayers show differences from superconductors, as evidenced by the current study. Hence, the existence of $\pi$-walls in bulk Sm-$A^\ast$ samples and their comparison to twist-grain-boundary phases, which are analogous to Abrikosov phases, are subject to further examination.

\section*{Acknowledgements}

We thank E. Barry, T. Gibaud, Z. Dogic, R. A. Pelcovits, H. Tu, and M. J. Zakhary for fruitful discussions. This work was supported by the NSF through MRSEC Grant no. 0820492.

\end{document}